\documentclass[aip,jmp,graphicx,reprint]{revtex4-1}
\draft
\usepackage{dcolumn}
\usepackage{graphicx}
\usepackage{subfigure}
\usepackage{epstopdf}
\usepackage{color}
\usepackage{amsmath}

\newcommand*{\citen}[1]{%
  \begingroup
    \romannumeral-`\x 
    \setcitestyle{numbers}%
    \cite{#1}%
  \endgroup   
}
\begin{document}

\title{Self-aligning concave relativistic plasma mirror with adjustable focus}

\author{Hai-En Tsai}
 \affiliation{Physics Department, The University of Texas at Austin, Austin, Texas 78712, USA} 
 
  \author{Alexey V. Arefiev}
  \affiliation{Institute for Fusion Studies, The University of Texas at Austin, Austin, Texas 78712, USA} 
  
  \author{Joseph M. Shaw}
  \affiliation{Physics Department, The University of Texas at Austin, Austin, Texas 78712, USA}

 \author{David J. Stark}
   \affiliation{Physics Department, The University of Texas at Austin, Austin, Texas 78712, USA} 
   \affiliation{Institute for Fusion Studies, The University of Texas at Austin, Austin, Texas 78712, USA} 
 
  \author{Xiaoming Wang}
 \affiliation{Physics Department, The University of Texas at Austin, Austin, Texas 78712, USA}

 \author{Rafal Zgadzaj}
 \affiliation{Physics Department, The University of Texas at Austin, Austin, Texas 78712, USA}

 \author{M. C. Downer}
  \email{downer@physics.utexas.edu}
  \affiliation{Physics Department, The University of Texas at Austin, Austin, Texas 78712, USA}
  
\date{\today}

\date{\today}

\begin{abstract}

We report an experimental-computational study of the optical properties of plasma mirrors (PMs) at the incident laser frequency when irradiated directly at relativistic intensity ($10^{18} < I_0 < 10^{19}$ W/cm$^2$) by near-normally incident ($4^\circ$), high-contrast, 30 fs, 800 nm laser pulses.  We find that such relativistic PMs are highly reflective ($0.6 - 0.8$), and focus a significant fraction of reflected light to intensity as large as $\sim 10I_0$ at distance $f$ as small $\sim 25 \mu$m from the PM, provided that pre-pulses do not exceed $10^{14}$ W/cm$^2$ prior to $\sim 20$ ps before arrival of the main pulse peak.  Particle-in-cell simulations show that focusing results from denting of the reflecting surface by light pressure combined with relativistic transparency, and that reflectivity and $f$ can be adjusted by controlling pre-plasma length $L$ over the range $0.5 \alt L \alt 3 \mu$m.  Pump-probe reflectivity measurements show the PM's focusing properties evolve on a ps time scale.   

\end{abstract}

\maketitle


\section{Introduction}

Plasma mirrors (PMs) have become standard tools for improving temporal contrast\cite{Kapteyn91,Ziener03,Doumy04,Geissel11,Backus93} and spatial profile \cite{Gold94} of intense, ultrashort laser pulses.   In this application, an ultrashort pulse is focused onto a solid\cite{Kapteyn91,Ziener03,Doumy04,Geissel11} or liquid \cite{Backus93} PM target to peak intensity in the range\cite{Kapteyn91,Ziener03,Doumy04,Geissel11,Gold94,Backus93,Thaury07} $10^{15} < I_0 < 10^{17} $ W/cm$^{2}$ at which only the intense core of the pulse generates plasma above the critical density, and thus reflects efficiently, while unwanted, less intense prepulses and spatial wings are transmitted.  A pulse thus ÒcleanedÓ by reflection from one or more primary PMs can subsequently be focused to relativistic intensity ($I \agt 10^{18} $ W/cm$^{2}$ for visible or near infrared wavelengths), where its interaction with a secondary PM can be studied without severely pre-expanding its surface.\cite{Thaury07,Vincenti14}  Such relativistic PMs generate high-order harmonics in the form of attosecond bursts,\cite{Thaury07,Vincenti14} and even focus those harmonics\cite{Vincenti14} when the light pressure of the incident pulse dents the PM's reflecting critical surface.  However, attempts to focus pulses to relativistic intensity at the primary PM often result in premature hydrodynamic surface expansion and premature generation of overdense plasma.\cite{Vincenti14,Adumi04}  

Recently new applications have emerged that demand high reflectivity and/or focusing at the \emph{fundamental} frequency \emph{directly} from a primary PM for light intensities in the range $10^{17} < I_0 < 10^{19}$ W/cm$^{2}$. These include self-aligned retro-reflection of the transmitted drive pulse of a laser-plasma accelerator (LPA) onto trailing electrons to produce bright Compton backscatter x-rays,\cite{Phuoc12, Tsai15} and coupling such a drive pulse over a short distance into the second or subsequent stage of a multi-stage LPA.\cite{Steinke16,Shaw16}  In addition, even for studies primarily devoted to highly nonlinear laser-PM processes such as high-order harmonic generation,\cite{Thaury07,Vincenti14,Borot11,Tilborg13} vacuum heating,\cite{Brunel87,Grimes99} hole-boring,\cite{Wilks92} or relativistic transparency,\cite{Palan12} quantitative characterization of the reflected main pulse is a valuable diagnostic of its relativistic interaction with the target.\cite{Schumacher11} 

Here we present an experimental-computational study of the optical properties of an initially planar PM at the fundamental frequency when excited directly at relativistic intensity.  A small portion of these measurements and simulations was published previously to support a study of Compton backscatter x-ray generation,\cite{Tsai15} and are included here for completeness.  Here, however, we greatly expand upon the previously reported measurements and simulations.  Specifically, we present new near- and far-field measurements of the reflected beam's focusing properties, pump-probe measurements of the relativistic PM's ultrafast response, and explicit characterization of pre-pulses and pre-plasma, along with new 2D PIC simulations for a wide range of $I_0$, pre-plasma lengths $L$ and time delays $\Delta t$ after excitation, and a 3D simulation of relativistic PM focusing.  These expanded results reveal two key properties of these unique relativistic optical elements.  First, their local transient reflectivity reaches values as high as $80 \%$ at $I_0$ up to $6 \times 10^{18}$ W/cm$^2$, provided the incoming pulse impinges at near-normal incidence and the intensity of its pre-pedestal does not exceed $10^{14}$ W/cm$^2$ prior to $\sim 20$ ps before its peak.  We controlled the pre-pedestal with optical elements inside the laser system, rather than upstream PMs, enabling the full amplified pulse energy to be focused onto the PM.  Second, the relativistic PM can \emph{focus} the reflected fundamental pulse with focal length $f$ as short as $25 \mu$m \emph{while reflecting it efficiently}, resulting in focused intensity $I_{\rm focus}$ as high as $10I_0$.   Previous work inferred similar $f$ from measurements of the far-field divergence of high-order harmonics from a relativistic PM,\cite{Vincenti14} but did not characterize the reflected fundamental pulse.   Here we \emph{directly} image the transverse profile of the reflected, focused \emph{fundamental} pulse both near to, and far from, the PM.  We find that measured far-field divergence does not accurately predict focused profile near the PM surface, a consequence of spherical aberration of the concave relativistic PM surface.   
Tighter focusing may be achievable by fabricating PMs with pre-curved substrates.\cite{Nakat10}  However, the present results show that light-induced concavity is alone sufficient to focus within tens of microns and to enhance intensity ten-fold using inexpensive planar foils.  Moreover, adjustment of intensity and pre-pedestal of the incident pulse can control the PM's focusing properties. 

Section II describes the experimental setup for four types of PM measurements:  time- and space-resolved reflectivity, time-integrated and space-resolved self-reflectivity, PM focusing, and pre-plasma length measurement.  Section III presents corresponding results of each measurement.  Section IV presents particle-in-cell (PIC) simulations, and discusses experimental and simulation results.  Section V presents our conclusions.


\section{Experimental Setup} 
\index{Experimental procedure%
@\emph{Experimental procedure}}%

Experiments used the 30 terawatt Ti:sapphire University of Texas Tabletop Terawatt (UT$^{3}$) laser system, which delivered pulses of wavelength $\lambda = 810$ nm, duration $\tau_{\rm FWHM} = 30$ fs and energy up to 1 J to the PM.  A temporal pulse cleaner based on cross-polarized wave (XPW) generation in barium fluoride (BaF$_2$),\cite{Jullien05} located prior to the main amplifiers of the UT$^{3}$ system, controlled the fully amplified pulse's peak-to-background contrast ratio, a critical parameter in performance of a PM irradiated at relativistic intensity.  

\begin{figure}[htb]
\centering
\includegraphics[trim={0 0 0 0.5cm}, width=0.38\textwidth]{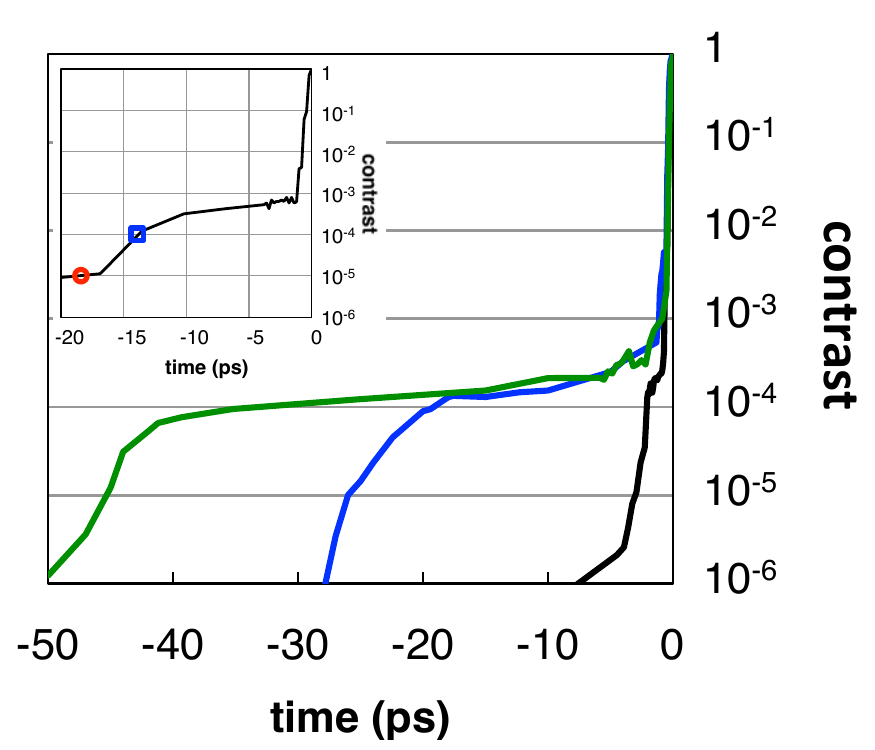} 
\caption{Scanning third-order autocorrelation traces of three different temporal pre-pulse pedestals achieved by varying the position of a BaF$_2$ crystal relative to beam focus within an XPW pulse cleaner. Inset:  pedestal profile that optimized PM performance at relativistic peak intensity.  Blue square (red circle) indicates estimated onset\cite{Adumi04} ($10^{14}$ W/cm$^2$) of pre-plasma formation for peak intensity $10^{18}$ ($10^{19}$) W/cm$^2$.  }
\label{contrast}
\end{figure}

The main panel of Fig.~\ref{contrast} shows three third-order autocorrelation\cite{Luan93} measurements of a pre-pulse pedestal of the fully amplified UT$^{3}$ pulse that was $\sim 10^4\times$ less intense than the peak.  Onset of this pedestal was varied from $40$ to $\sim 1$ ps before the peak by adjusting intensity in the BaF$_2$ crystal in the XPW pulse cleaner.  With peak intensity $I_0 \agt10^{18}$ W/cm$^2$ on target, this pedestal became sufficiently intense ($\agt10^{14}$ W/cm$^2$) to pre-expand the PM surface.\cite{Adumi04}  Adjustment of the pedestal duration enabled control of the pre-plasma scale length $L$, which in turn strongly influenced PM reflectivity, focal length and optical quality.  
The inset of Fig.~\ref{contrast} shows a pedestal profile that yielded the most reflective and strongly focusing PM at relativistic intensity of those investigated, and that proved stable in day-to-day operation of the UT$^{3}$ system.  We therefore used it for all measurements presented below.  We can estimate $L$ from this pre-pulse profile and previous measurements\cite{Adumi04} of plasma expansion speed $C_s$ of fs-laser-excited targets.  Adumi \textit{et al.}\cite{Adumi04} measured $C_s \approx 100 \mu$m/ns for aluminum targets excited by an ultrashort pre-pulse of intensity $10^{15}$ W/cm$^2$.
We choose this intensity to take into account plasma heating by pedestal light following the time $t$ at which the pedestal reaches the pre-plasma``trigger" threshold $\sim 10^{14}$ W/cm$^2$, marked in Fig.~\ref{contrast} (inset) by a blue square (red circle) for peak intensity $I_0 = 10^{18}$ ($10^{19}$) W/cm$^2$.  Using this $C_s$ value for our SiO$_2$ targets yields $L \approx C_s|t| \approx 1.5 (2) \mu$m for $I_0 = 10^{18}$ ($10^{19}$) W/cm$^2$.    For $I_0 = 10^{17}$ W/cm$^2$, the trigger threshold is reached at $t \approx -2$ ps ($10^{-3}$ contrast), yielding $L \approx 0.2 \mu$m.  


\begin{figure*}[htb]
\centering
\includegraphics[width=0.7\textwidth]{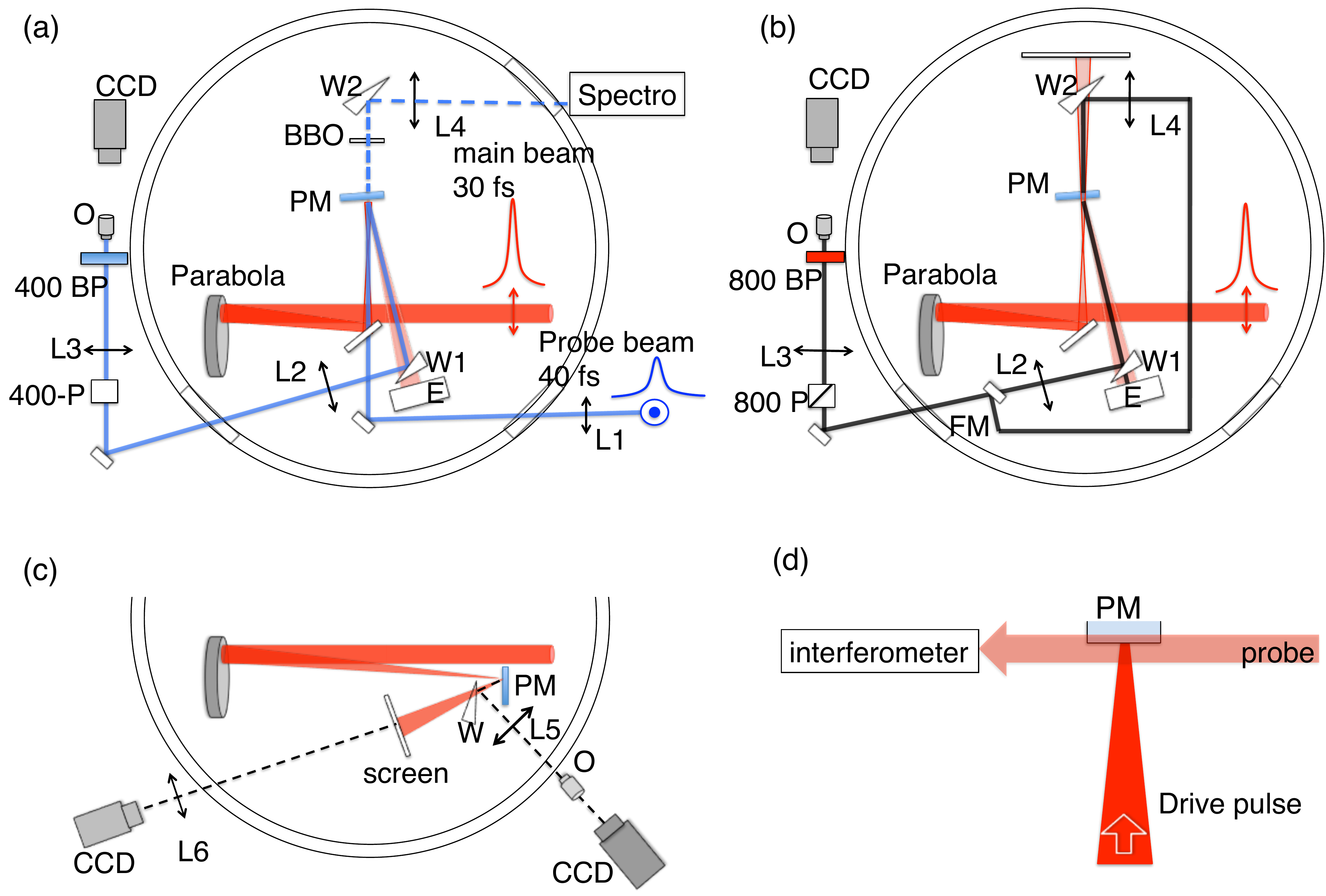}
\caption{
Schematic experimental setup for four types of PM measurements. \textbf{(a)} Time- and space-resolved PM reflectivity: blue (red) line shows path of 400-nm probe (800-nm main) beam. \textbf{(b)} Time-integrated, space-resolved PM reflectivity:  black line shows optical path of reflected and transmitted 800 nm beam through the imaging system. \textbf{(c)} Relativistic PM focusing:  black dashed lines show paths of reflected 800 nm beam through two separate imaging systems that measured far-field beam divergence and focused beam profile near PM surface. \textbf{(d)} Pre-plasma characterization via transverse interferometery.  Key to abbreviations:  BP = bandpass filter, CCD = charge-coupled device camera, E = energy meter, FM = flip mirror, L = lens, O = objective lens ($20\times$), P = polarizer, PM = plasma mirror, Spectro = spectrometer, W = wedge.  Focal lengths of lenses are 1.0 m (L1), 0.375 m (L2-L4), 0.2 m (L5), 0.1 m (L6).  Red double arrow (blue circled dot) indicates pump (probe) polarization. }
\label{figure_PM_setup}
\end{figure*}  

Fig.~\ref{figure_PM_setup} shows the target chamber setup for four types of PM measurements. For time-resolved PM reflectivity measurements [Fig.\ref{figure_PM_setup}(a)], a probe pulse was split from the main pump pulse, frequency-doubled, and propagated to the target through a variable delay line (not shown).  
The main pulse entered the chamber with a rounded top-hat transverse spatial profile with $99\%$ of its energy within a $\sim 5$cm diameter.  A gold-coated off-axis parabolic mirror ($f = 60$ cm) focused it with $f/12$ onto the PM target with incidence angle $4^\circ$, $p$-polarization, Strehl ratio $>0.5$.  Direct imaging of the vacuum focus yielded an intensity profile of diameter $w_{\rm FWHM} = 11 \pm 1 \mu$m, which for a Gaussian profile implies 1/e field profile radius $w_0 = w_{\rm FWHM}/\sqrt{2\ln2} = 10 \pm 1 \mu$m and peak intensity up to $\sim10^{19} \rm\ W/cm^2$ ($a_0 \approx 2$).  The uncertainties indicate rms shot-to-shot fluctuations and spatial asymmetry.  Here the dimensionless laser strength parameter $a_0 = eE_0/mc\omega$ is the ratio of the momentum $eE_0/\omega$ per optical cycle that the laser field $E_0$ (frequency $\omega$) imparts to an electron (charge $e$, mass $m$) to $mc$; thus $a_0 \geq 1$ defines the relativistic excitation regime.

Lens L1 ($f =1$ m) focused the 50 fs, 5-mm-diameter incident probe to $w_{\rm FWHM} \sim 100 \mu$m onto the PM surface colinearly with, and polarized orthogonally to, the pump, and overlapping the pumped region.  After the probe reflected from the PM, a second reflection from wedge W1 directed a portion of it to imaging system L2-L3-O. 
Single-shot calorimeter (E) measured probe energy transmitted through the wedge, and was used to calibrate absolute reflectivity ($5\%$) of the \emph{unpumped} PM.  
With the probe blocked, calorimeter E also measured space-time-integrated self-reflectivity of the main 800 nm pulse with $f/3$ acceptance cone.

For space-resolved PM reflectivity measurements, lenses L2, L3, and O relay-imaged the reflected probe [Fig.\ref{figure_PM_setup}(a)] or main beam [Fig.\ref{figure_PM_setup}(b)] from the PM surface to a 12-bit CCD camera on separate shots through 400 or 800 nm bandpass filters and orthogonally-oriented 400 or 800 nm polarization analyzers.  The longitudinal position of L2 was adjusted to optimize imaging.  The $f/12$ acceptance cone of L2 sufficed to collect nearly all reflected light at sub-relativistic pump intensity.  At higher pump intensity, however, some reflected light diverged beyond its perimeter, as discussed in Sections III and IV. 
To image the profile of the focused main beam \emph{incident} on the PM surface, the PM was removed, and wedge W2 and lens L4 (identical to W1 and L2, respectively) were inserted along with flip mirror FM [see Fig.~\ref{figure_PM_setup}(b)]. The longitudinal position of L4 was adjusted to image the incident profile at the PM plane to the same CCD, with the same magnification, as the reflected profile.  The ratio of these images yielded absolute space-resolved self-reflectivity.
In the pump-probe configuration [Fig.\ref{figure_PM_setup}(a)], ratios of probe images with and without the main beam at various pump-probe delays similarly yielded absolute time- and space-resolved reflectivity. 
 

\begin{figure*}[htb]
\centering
\includegraphics[width=0.7\textwidth]{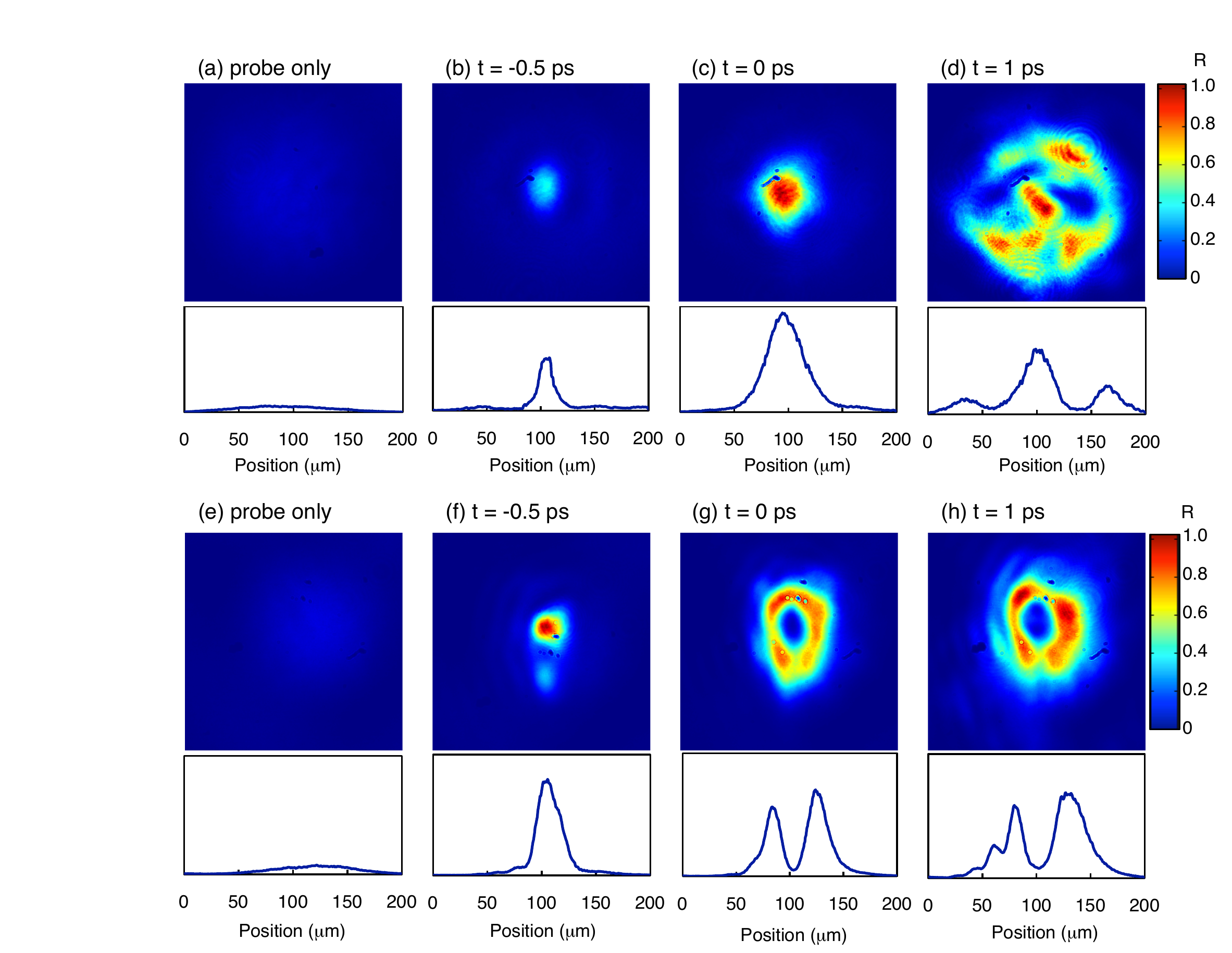}
\caption{Time-resolved spatial intensity profiles of 50 fs, 400 nm probe pulse as it reflects from PM without a pump pulse \textbf{(a)}, \textbf{(e)}, and at various time delays $t$ with respect to the peak of a 30 fs, 800 nm pump pulse of intensity $10^{17}$ W/cm$^2$ (top row) or $5\times10^{18}$ W/cm$^2$ (bottom row):   \textbf{(b)}, \textbf{(f)} $t = -0.5$ ps; \textbf{(c)}, \textbf{(g)} $t = 0$ ps; \textbf{(d)}, \textbf{(h)} $t = 1$ ps. }
\label{figure_PM_prb}
\end{figure*}
\normalsize

To image the profile of reflected light immediately in front of the PM [Fig. \ref{figure_PM_setup}(c)], 
wedge W, located 10 cm from the PM, directed $\sim1\%$ of reflected energy through lens L5 ($f=20$ cm, $f/5$), a chamber window, and objective lens O onto a 12-bit CCD camera. The longitudinal position of O was adjusted to image reflected beam profiles immediately in front of the PM surface to characterize near-field focusing by its dented surface. The L5-O microscope imaged transverse profiles with $\sim2\mu$m resolution, and determined image plane position $x$ above the PM surface to within $\pm 5\mu$m. 

The part of the beam transmitted through wedge W was projected onto a white screen, which lens L6 ($f = 10$ cm) imaged onto another 12-bit CCD camera.  
Changes in the far-field divergence of the reflected beam were then correlated with changes in its near-field focusing. 

A transverse interferometer [Fig.~\ref{figure_PM_setup}(d)] was used to characterize the pre-plasma density profile $n_e(x)$ at distance $x$ above the PM.  It is described further in the Supplementary Material.  

\begin{figure*}[t]
\centering
\includegraphics[width=0.9\textwidth]{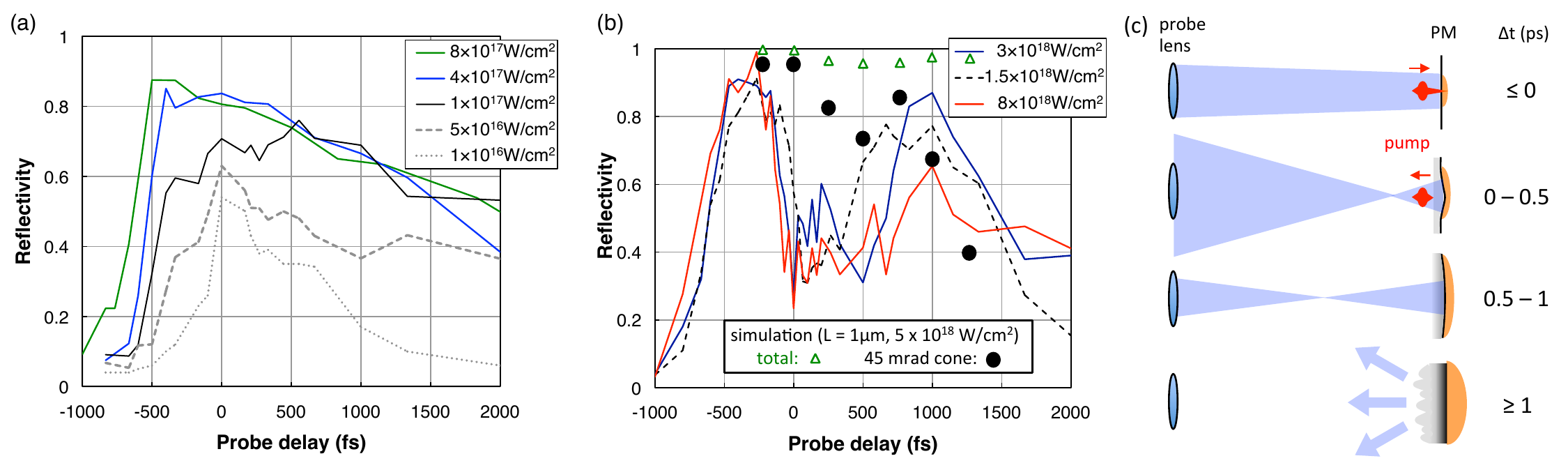}
\caption{Measured reflectivity of 50 fs, 400 nm probe pulse from center of photo-excited spot of PM  vs.~time delay $\Delta t$ after excitation, for pump intensities (a) $10^{16} \leq I_0 \leq 8\times 10^{17}$ W/cm$^2$ and (b) $I_0 > 10^{18}$ W/cm$^2$.  Filled black circles:  2D simulation of fraction of probe energy reflected into a 45 mrad cone at seven $\Delta t$ after exciting PM ($L = 2\mu$m) at $I_0 = 5 \times 10^{18}$ W/cm$^2$; green triangles:  total probe reflectivity vs.~$\Delta t$. (c) Conceptual picture of evolving PM and probe reflectivity during and after pump excitation.  Light blue:  reflected probe trajectory; light grey:  pre-plasma density profile; orange:  overdense plasma. }
\label{figure_PM_prbplot}
\end{figure*}


\section{Experimental Results}


\subsection{\label{sec:level2} Pump-probe PM reflectivity measurements}
Figure~\ref{figure_PM_prb} presents images of the probe pulse spatial profile as it reflected from the PM at time delays $t=-0.5$ (b, f), 0 (c, g) and 1 ps (d, h) from the peak of the main pulse, which was focused to peak intensity $10^{17}$ W/cm$^2$ (top row) or $5 \times10^{18}$ W/cm$^2$ (bottom row).  Panels (a, e) show probe reflectivity from the un-pumped PM.   
For both pump intensities, PM reflectivity rose already by $t = -0.5$ ps (b, f).  However, the reflected profiles evolved differently over the interval $-0.5 < t < 1$ ps.  At $10^{17}$ W/cm$^2$, probe reflectivity rose initially (b) to $50\%$ within a $10 \mu$m diameter region, an imprint of the focused pump pulse profile.  By $t = 0$ ps (c), this region grew to $50 \mu$m diameter with $70\%$ reflectivity.  Finally, at $t = 1$ ps (d), the reflective region grew to the diameter of the probe ($\sim100 \mu$m), with modulations attributable to uneven plasma expansion.  In contrast, at $5 \times10^{18}$ W/cm$^2$, central reflectivity was already $80\%$ at $t = -0.5$ ps (f), and the reflective region was $20 \mu$m in diameter.   By $t = 0$ ps (g), the reflective region expanded, but a hole appeared in the center, forming a reflective ring with $70\%$ reflectivity.  The low central reflectivity was still visible at $t = 1$ps delay (h), but recovered shortly thereafter.

Figure~\ref{figure_PM_prbplot} shows time-resolved reflectivity at the central, directly photo-excited spot on the probe profile for eight different pump intensities. 
For pump intensities $I_0 < 10^{18} $W/cm$^{2}$ [Fig.~\ref{figure_PM_prbplot}(a)], probe reflectivity reached a maximum at times in the range $-0.5 < t < 0$ ps, then decayed monotonically over several ps.  For $I_0 > 10^{18} $W/cm$^{2}$ [Fig.~\ref{figure_PM_prbplot}(b)], probe reflectivity rose to a first maximum at $t \sim -0.5$ ps, then dropped sharply near $t = 0$ ps, a temporal ``hole" that corresponds to the hole in the probe spatial profile shown in Fig.~\ref{figure_PM_prb}(g).  Center reflectivity then recovered, reaching a second maximum at $t \agt 1$ ps before decaying again over several ps as the plasma expanded further.  This transient spatial-temporal hole results from ultrafast changes in PM curvature, as shown in Fig.~\ref{figure_PM_prbplot}(c) and discussed in Sec.~IV.

At $I_0 = 8\times10^{18}$ W/cm$^2$, PM reflectivity began rising at  $t \sim -1$ ps [Fig.~\ref{figure_PM_prbplot}(b)].  From the temporal contrast at $t = -1$ ps in the inset of Fig.~\ref{contrast} ($\sim5\times10^{-4}$), this corresponds to threshold intensity $\sim5\times10^{14}$ W/cm$^2$ (fluence $\sim10$ J/cm$^2$), in good agreement with thresholds for high reflectivity reported in previous PM studies.\cite{Ziener03,Doumy04} For $1 \alt I_0 \alt 8\times 10^{18}$ W/cm$^2$, this threshold occurs only $\sim 0.1$ps later because of the steep rising edge of the main pulse (Fig.~\ref{contrast}, inset), consistent with the onset of reflectivity rise at these intensities shown in Fig.~\ref{figure_PM_prbplot}(b).


\subsection{Time-integrated PM reflectivity measurements}
\index{Time-integrated reflectivity of the PM%
@\emph{Time-integrated reflectivity of the PM}}%

\begin{figure}[htb]
\centering
\includegraphics[trim={0.5cm 0.5cm 0 1.7cm},width=0.40\textwidth]{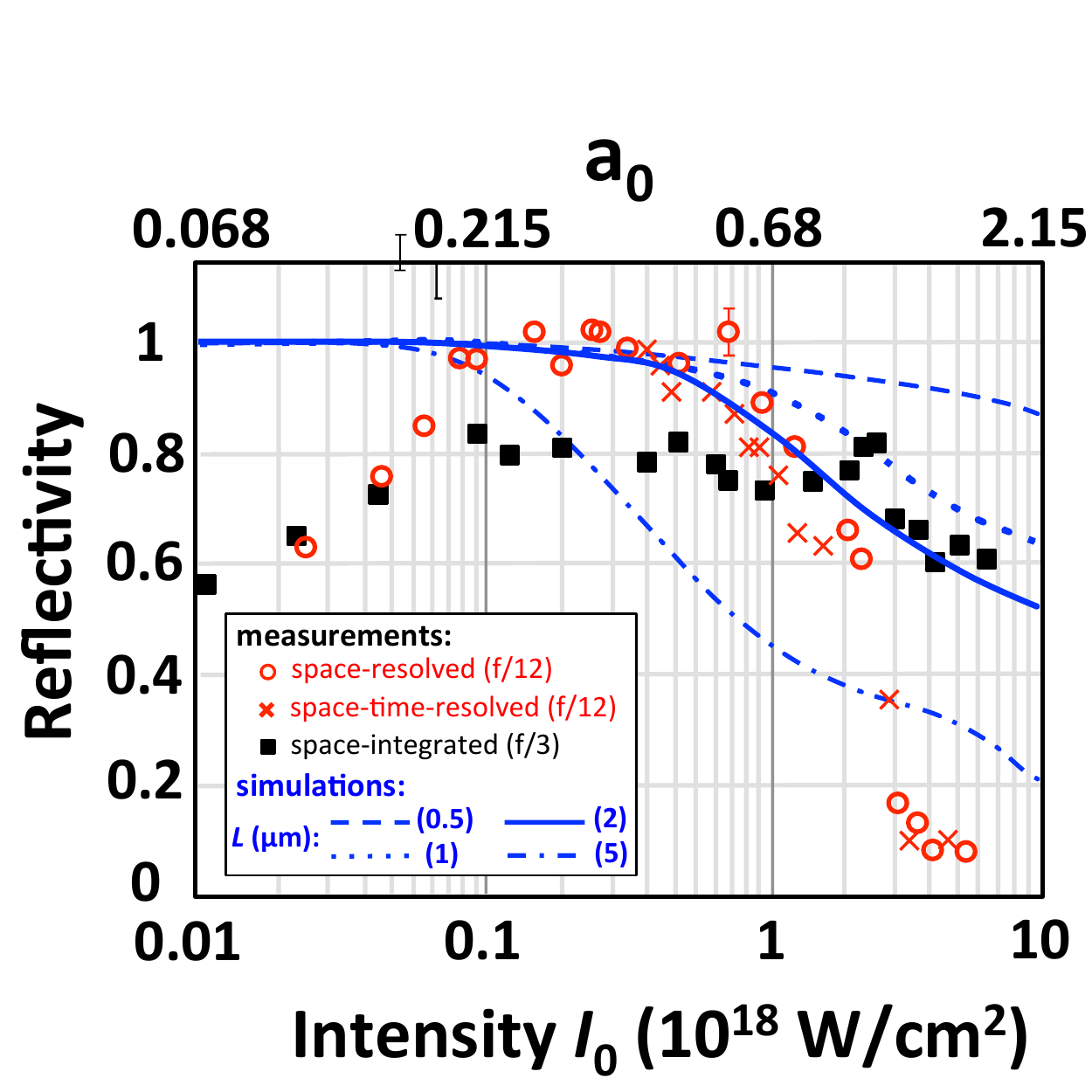}
\caption{Reflectivity of PM vs.~incident pump intensity $I_0$. Red circles (crosses): pump self-reflectivity (probe reflectivity at $t=0$) at intense center of pump profile, measured through telescope with $f/12$ collection cone; black squares:  integrated self-reflectivity of entire pump pulse, measured with calorimeter with $f/3$ collection cone; blue curves: simulated net PM reflectivity, assuming pre-plasma layer of thickness $L = 0.5 \mu$m (dashed), 1 $\mu$m (dotted), $2 \mu$m (solid), or $5 \mu$m (dot-dashed). The time-integrated data (filled black squares, red circles) for $I_0 > 10^{17}$ W/cm$^2$ and a preliminary version of the simulations for $L = 0.5, 1.0 \mu$m (blue dashed, dotted curves) were included in Fig.~6 of Ref.~11, and are repeated here for completeness.
}
\label{figure_PM_reflectplot}
\end{figure}

Fig.~\ref{figure_PM_reflectplot} presents results of two measurements of intensity-dependent self-reflectivity of the 800 nm pump pulse from the PM:  (\textit{i}) space-resolved reflectivity of the intense center of the reflected pump profile (red circles), measured using the L2-L3-O telescope with $f/12$ collection cone [Fig. \ref{figure_PM_setup}(b)]; (\textit{ii}) space-integrated reflectivity of the entire pump profile (filled black squares), measured using calorimeter E with $f/3$ collection cone [Fig.~\ref{figure_PM_setup}(b)].  For comparison, reflectivity at $t=0$ at the pump-irradiated center of the \emph{probe} profile, also measured using the L2-L3-O telescope, is plotted (red crosses) for $I_0 > 3 \times 10^{17}$ W/cm$^2$.  The last measurements agreed closely with the space-resolved self-reflectivity measurements, even though the probe was frequency-doubled.  Below we refer to these two sets of measurements jointly as ``space-resolved reflectivity".  The blue curves are simulation results, discussed in Sec. IV.


\normalsize

For pump intensities in the range $10^{16}$ to $10^{17}$ W/cm$^2$, space-integrated and space-resolved reflectivity measurements both yielded $60$ to $80\%$ reflectivity.  From $10^{17}$ to $\sim10^{18}$ W/cm$^{2}$, space-integrated calorimeter measurements (black squares) yielded $\sim80\%$ reflectivity, whereas as space-resolved measurements yielded 90 to 100\% reflectivity.  The lower space-integrated value in this range is straightforwardly explained if $\sim20\%$ of the pump pulse energy lay outside the central focus and failed to reach a threshold intensity needed to create a highly reflective overdense plasma.\cite{Ziener03,Doumy04}  Space-resolved measurements, in contrast, pre-selected this highly reflective region.  For relativistic pump intensities $>10^{18}$ W/cm$^{2}$, on the other hand, the wide ($f/3$) cone calorimeter measurements (black squares) yielded much \emph{higher} reflectivity:   $\sim80\%$ up to $2\times10^{18}$ W/cm$^{2}$, dropping gradually to $\sim 60\%$ as intensity increased to $5 \times 10^{18}$ W/cm$^{2}$. In contrast, reflectivity from narrow ($f/12$) cone, space-resolved measurements (red circles, crosses) drop steeply to $< 10\%$ at $5 \times 10^{18}$ W/cm$^{2}$.   This discrepancy could be explained either by strong absorption localized in the intense center of the pulse profile, or by defocusing of the reflected pulse outside the $f/12$ collection cone due to curvature of the PM surface.  In Sec.~IV, we discuss these options in the context of other measurements, and simulations.



\begin{figure}[ht]
\centering
\includegraphics[trim={0.3cm 0 0 0.11cm},width=0.4\textwidth]{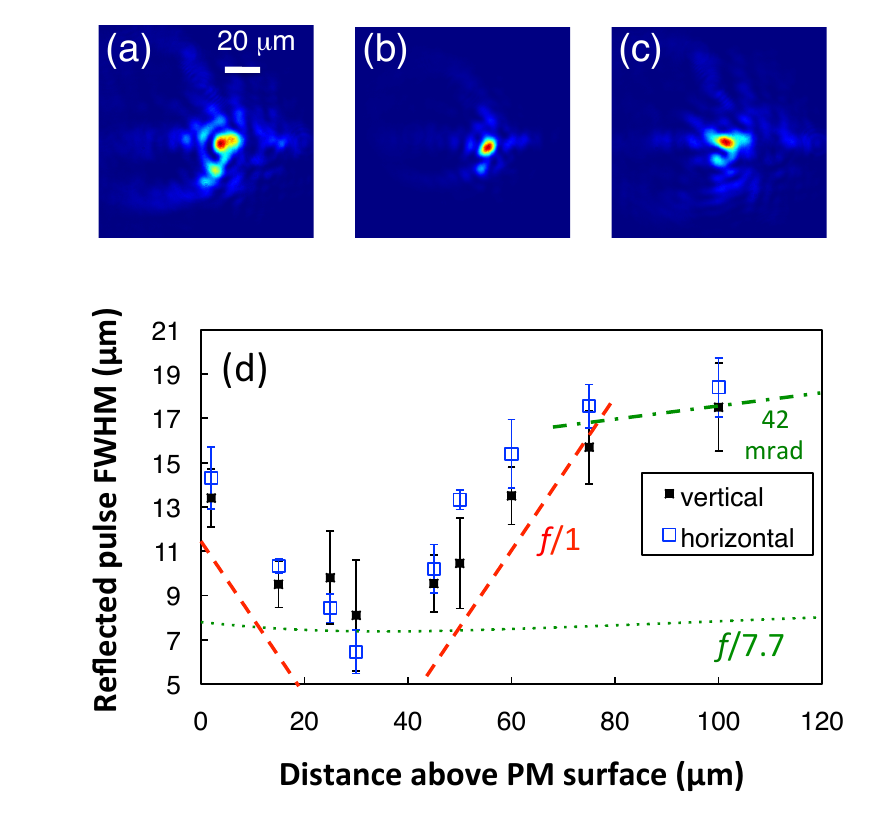}
\caption{Focusing of relativistically intense ($6\times10^{18}$ W/cm$^2$) laser pulse by concave PM. Top row: spatial profiles of reflected pulse at distances $x =$ \textbf{(a)} 0, \textbf{(b)} 30 and \textbf{(c)} 60$\mu$m above PM surface. \textbf{(d)} Plot of FWHM of reflected pulse along two orthogonal axes vs.~$x$.  Error bars indicate rms shot-to-shot fluctuations. Curves show $w_{\rm FWHM}(x)$ for various Gaussian beams:  $f/1$ focus with $w_{\rm FWHM}(x=0) = 11\mu$m at PM surface (red dashed); $f/7.7$ focus with $w_{\rm FWHM}(x=30\mu{\rm m}) = 7\mu$m (green dotted); 42 mad divergence, fit to data points at $x=75, 100 \mu$m (green dot-dashed).  The last two curves correspond to measured far-field divergence at $6\times10^{18}$ W/cm$^2$.}
\label{figure_PM_2ndfocus}
\end{figure}

\subsection{PM focusing measurements}
\index{Relativistic denting effect on a PM: measurement%
@\emph{Relativistic denting effect on a PM: measurement}}%

Panels (a)-(c) of Fig.~\ref{figure_PM_2ndfocus} show transverse spatial profiles of the 800 nm pump pulse, imaged through the L5-O telescope in Fig.~\ref{figure_PM_setup}(c), at distances (a) 0, (b) 30 and (c) $60 \mu$m above the PM surface after reflecting from it at intensity $5\times10^{18}$ W/cm$^2$.  The profiles, obtained on separate shots with different longitudinal positions of objective O, have average FWHM (a) $14\mu$m, (b) $7\mu $m and (c) $18\mu $m, showing that the reflected pump converged to a focus at $x \sim 30 \mu$m, then diverged at larger $x$.   Fig.~\ref{figure_PM_2ndfocus}(d) plots the results of a larger set of FWHM measurements over the range $0 < x < 100 \mu$m, distinguishing FWHM orthogonal (``vertical") and parallel (``horizontal") to the plane of incidence.  The convergence ($0 < x < 30 \mu$m) and divergence ($30 < x < 60 \mu$m) of the near-field focus roughly match those of an $f/1$-focused Gaussian beam of diameter $w_{\rm FWHM}(x=0) = 11 \mu$m at the PM surface (red dashed curve).  Measured $w_{\rm FWHM}(x)$, however, are larger than those of such an ideal Gaussian beam because of limited imaging resolution, uncertainty of the object plane of the L5-O imaging system, and non-Gaussian features of the focusing beam, as discussed in Sec. IV.  


\begin{figure}[ht]
\centering
\includegraphics[trim={0.3cm 0 0 0.3cm},width=0.45\textwidth]{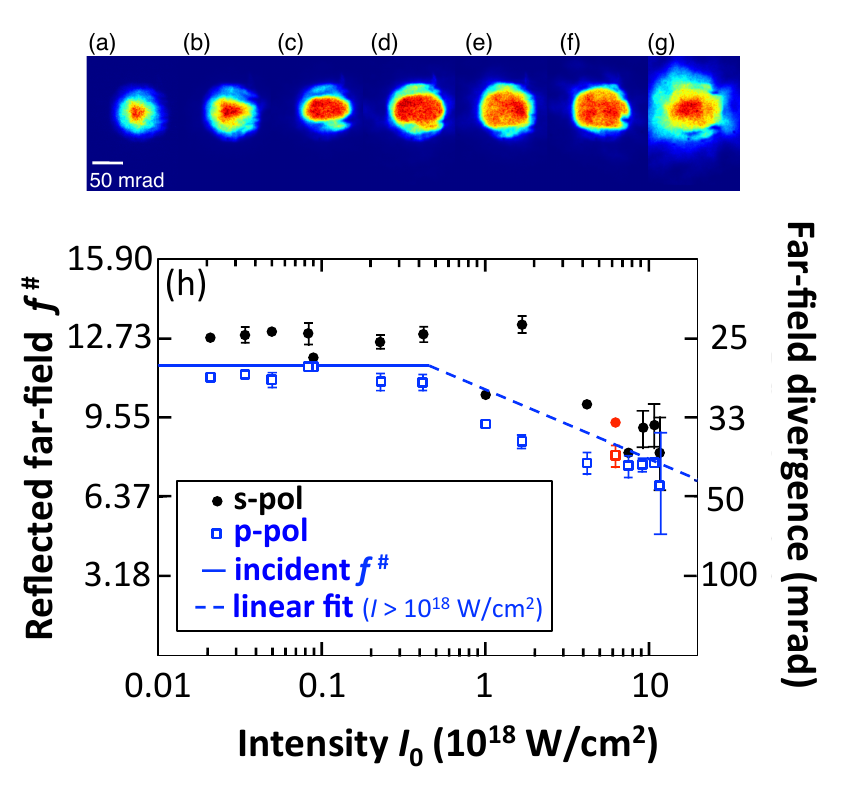}
\caption{Top row:  Far-field spatial profiles of pump pulse after reflecting from PM at intensity \textbf{(a}) $.084$, \textbf{(b)} $0.9$, \textbf{(c)} $1.7$, \textbf{(d)} $4.0$, \textbf{(e)} $6.0$, \textbf{(f)} $7.5$, and \textbf{(g)} $12\times10^{18}$ W/cm$^2$.  Panel \textbf{(h)} plots far-field $f^\#$ and divergence $\theta = 1/\pi f^\#$ of reflected s- and p-polarized pulses. Red data points correspond to conditions of Fig.~6.  Error bars denote rms shot-to-shot fluctuations.  Blue solid line:  $f^\#$ of incident beam; blue dashed line:  linear fit to measured divergence for $I > 10^{18}$ W/cm$^2$. 
}  \normalsize 
\label{figure_PM_DIV}
\end{figure}

Figure \ref{figure_PM_DIV} presents corresponding measurements of far-field divergence of pump pulses reflected from the PM.  Images (a)-(g) show transverse profiles of reflected pulses far from the PM as $I_0$ increased.  Vertical (horizontal) axis lies orthogonal (parallel) to the plane of incidence.    Reflected profiles coincided with low-intensity profiles for $I_0$ up to $\sim 7\times10^{17}$ W/cm$^2$ [Fig.~\ref{figure_PM_DIV}(a)],  started expanding as $I_0$ approached $10^{18}$ W/cm$^2$ [Fig.\ref{figure_PM_DIV}(b)], and continued at higher $I_0$ [Fig.~\ref{figure_PM_DIV}(c)-(g)].  

Fig.~\ref{figure_PM_DIV}(h) plots the far-field $f^\#$ of the reflected beam --- \textit{i.e.} the ratio of PM-screen distance to vertical/horizontal diameters of an ellipse enclosing $99\%$ of the energy of far-field profile --- vs.~$I_0$.  For $I_0$ up to $\sim 7\times10^{17}$ W/cm$^2$, $f^\#$ remained constant at $\sim12$ (divergence half-angle $\theta\sim27$ mrad) --- equivalent to the incident focus (Fig.~\ref{figure_PM_DIV}(h), solid blue line) --- then dropped to $\sim7$ ($\theta\sim45$ mrad) as $I_0$ approached $10^{19}$ W/cm$^2$.  The measured far-field $f/7.7$ ($\theta = 42$ mrad) at $I_0=6\times10^{18}$ W/cm$^2$ (red data points in Fig.~\ref{figure_PM_DIV}(h)) would correspond to Gaussian beam waist $w_{\rm FWHM} = \sqrt{2 ln 2} (f^\#\lambda) \approx 7\mu$m, coincidentally equal to the measured waist $w_{\rm FWHM}(x=30\mu{\rm m})$ in Fig.~\ref{figure_PM_2ndfocus}(d).  However, this apparent agreement is fortuitous because an $f/7.7$ beam would have diverged much more gradually  in the near field than observed (Fig.~\ref{figure_PM_2ndfocus}(d), dotted green curve).  The more gradual beam expansion observed for $x > 80 \mu$m (Fig.~\ref{figure_PM_2ndfocus}(d)) may indicate a transition from near- to far-field divergence.  Evidently at relativistic intensity the reflected near- and far-field focus profiles shown in Figs.~\ref{figure_PM_2ndfocus} and \ref{figure_PM_DIV}, respectively, originate from separate components of the reflected beam.  The properties of one cannot be inferred from the other.  Taken together, these measurements demonstrate that the PM becomes a concave, albeit spherically aberrated, focusing mirror with intensity-dependent focal lengths when excited in the range $10^{18}$ to $10^{19}$ W/cm$^2$.   


\subsection{PM preplasma characterization}
\index{Preplasma characterization: measurement%
@\emph{Preplasma characterization: measurement}}%

Transverse interferometry measurements used a probe pulse propagating along the PM surface [Fig.~\ref{figure_PM_setup}(d)] that crossed the path of the pump pulse 0.3 ps before its peak.  The interferometer was sensitive to plasmas of density $n_e \agt 10^{18}$ cm$^{-3}$ of scale length $L \agt 5 \mu$m.  However, no pre-plasma was detected up to $I_0 \sim 10^{19}$ W/cm$^2$).  This null result places an upper limit $L< 5 \mu$m on pre-plasma length. The Supplementary Material shows a transverse interferogram at $I_0 = 6 \times 10^{18}$ W/cm$^2$.


\section{Simulations and Discussion}
To help understand the experimental results in Figs.~\ref{figure_PM_prb}-\ref{figure_PM_DIV}, we performed simulations using the multi-dimensional, fully electromagnetic, relativistic PIC code EPOCH.\cite{EPOCH}  The simulations aimed to quantify how much the PM \emph{absorbed} and \emph{focused} the laser pulse as \emph{incident intensity $I_0$}, \emph{pre-plasma length $L$} and \emph{time delay} $\Delta t$ varied.  For 2D simulations, a $y$-polarized laser pulse with $\lambda = 800$nm and Gaussian transverse profile $a_0e^{-y^2/w_0^2}$ 
impinged on the PM along the $x$-axis with $w_0 =10 \mu$m, duration $\tau_{FWHM} = 30$ fs. The simulation box in the $(x,y)$-plane was $170 \times100 \mu$m, with $9600\times5000$ cells, respectively.  To simulate time-resolved reflectivity (Fig.~\ref{figure_PM_prbplot}),  frequency-doubled probe pulses with $w_0 =25 \mu$m, $\tau_{FWHM} = 30$ fs and $a_0 \sim .01$ co-propagated with the pump at selected $\Delta t$.  For 3D simulations, we used a fully Gaussian pulse $a_0 e^{-(y^2+z^2)/w_0^2}$ of the same $w_0$ and $\tau_{FWHM}$ in a $40\times 30\times 30\mu$m domain ($1700\times 300\times 300$ cells).   

To simplify the simulations, we assumed a pre-ionized PM with fixed pre-plasma of exponential form
\[    n_e(x)= 
\begin{cases}
    50n_{\rm{crit}}\exp[(x-x_0)/L],& \text{for } x\leq x_0\\
    50n_{\rm{crit}},              & \text{for } x> x_0.
\end{cases}
\]
This profile mimicked the effect of a pre-pulse on a PM with initially sharp boundary at $x=0$.  As the pre-plasma expanded to length $L$ in front of the target, the unperturbed boundary receded to $x_0$ to satisfy mass conservation.   Ionization levels were set to those that would be achieved by tunneling ionization in a laser field with intensity $>10^{17}$ W/cm$^2$. Inside the unperturbed part of the mirror ($x>x_0$), this led to $n_e=50n_{\rm{crit}}$, where  $n_{\rm{crit}}$ is the classical critical density for  $\lambda = 800 nm$ light.   No additional ionization took place during the simulations.   For 2D simulations, we used 15 Silicon (Z=12), 15 Oxygen (Z=6), and 50 electron macroparticles per cell.  For 3D simulations, we used
4 ions (of each species) and 10 electrons per cell.  We checked the validity of the latter resolution by rerunning 2D simulations with the same parameters, and confirming that the main results (reflected pulse amplitude, PM focal length) were unchanged.

\subsection{PM absorption}
\index{PM absorption%
@\emph{PM absorption}}%
 
 \subsubsection{Absorption mechanisms}
Absorption of laser energy by the PM is the only source of net reflectivity decrease.  At sub-relativistic intensity, the laser pulse loses energy mainly through target ionization and electron collisions in the resulting plasma. We did not simulate ionization losses here, since we assumed pre-ionized plasma, but we estimate them to be $< 1 \%$ by taking into account the ionization thresholds, density and volume of ionized target atoms.  At relativistic intensity, the pulse loses energy mainly by exciting longitudinal electron motion in an amount determined by the interaction length (see Ref.~\citen{arefiev2014} and references therein for a full discussion). We calculated net reflectivity by taking the ratio of the total energy in the reflected pulse to the total energy in the incoming laser pulse. 

\subsubsection{Simulated and measured absorption}
The blue curves in Fig.~\ref{figure_PM_reflectplot} show results of 2D PIC simulations of net reflectivity for pre-plasmas with $L = 0.5 \mu$m (dashed), $1$$\mu$m (dotted), $2$$\mu$m (solid) and $5 \mu$m (dot-dashed).    For $I_0  <  5\times10^{17}$ W/cm$^2$, the first three simulations yielded $\sim 100\%$ reflectivity.  This agrees with near unity reflectivity observed in space-resolved measurements for $10^{17} < I_0 < 5 \times 10^{17}$ W/cm$^2$ (Fig.~\ref{figure_PM_reflectplot}, red circles and crosses).  The growth in net reflectivity observed from $10^{16}$ to $10^{17}$ W/cm$^2$ can be attributed to the growing transverse size of the highly reflective plasma, an effect not included in the simulations since we assumed pre-ionized plasma.  Thus the measured $80\%$ spatially-integrated net reflectivity observed for $10^{17} < I_0 < 5 \times 10^{17}$ W/cm$^2$ (Fig.~\ref{figure_PM_reflectplot}, filled black squares) must be attributed to $20\%$ of the pulse energy residing in non-Gaussian wings or side lobes that fail to create an overdense plasma, also an un-simulated effect. 

For $I_0 \agt 5\times10^{17}$ W/cm$^2$, the simulations showed substantial absorption $A$ that increased monotonically with $I_0$ and $L$.  For the highest intensity ($5\times10^{18}$ W/cm$^2$) at which net reflectivity $R = 1-A$ was measured ($R\sim 0.65$), simulations for $L = 0.5, 1, 2$ and $5 \mu$m yielded $R = 0.91$, $0.68$, $0.59$ and $0.31$, respectively.  As expected,\cite{arefiev2014} $A$ was greater for longer pre-plasmas.  

Since the simulated reflectivity in Fig.~\ref{figure_PM_reflectplot} does not include effects of finite spot size or defocusing of reflected light that infect some of the reflectivity data, the simulations should be compared with those subsets of the data that best avoid these artifacts.  This choice depends on the $I_0$ range. Data in the range $10^{16} \alt I_0 \alt 10^{17}$ W/cm$^2$ should be neglected for this comparison, since the growing size of the overdense plasma within the focused laser spot dominates the observed reflectivity.  For $10^{17} \alt I_0 \alt 5\times10^{17}$ W/cm$^2$, space-resolved reflectivity measurements (Fig.~\ref{figure_PM_reflectplot}, red circles, crosses) best represent simulated reflectivity, since they avoid the low-$R$ contribution from the spatial wings of the incident pulse, yet defocusing of reflected light out of the f/12 collection cone is not yet strong.  For $I_0 \agt 5\times10^{17}$ W/cm$^2$, on the other hand, space-integrated reflectivity measurements (Fig.~\ref{figure_PM_reflectplot}, filled black squares) best represent simulated reflectivity, since the overdense plasma has grown to include the wings of the focused spot, while the wide $f/3$ collection cone avoids loss of defocused light.  

Using these criteria, the simulation with $L = 2$ $\mu$m agreed best with space-resolved reflectivity measurements for $10^{17} \alt I_0 \alt 5\times10^{17}$ W/cm$^2$, and with space-integrated reflectivity measurements for $I_0 \agt 5\times10^{17}$ W/cm$^2$.  This $L$ is also close to the value estimated for $10^{18} < I_0 < 10^{19}$ W/cm$^2$ from pre-pulse characterization (Fig.~\ref{contrast}, inset) and plasma expansion speed measurements\cite{Adumi04} (see Sec. II), and is consistent with interferometry measurement placing an upper limit $L < 5 \mu$m (see Sec. III.D).  Taken together, these three lines of evidence all independently point to $L \approx 2 \mu$m as the pre-plasma scale length that prevails for our PM measurements at relativistic intensity.  On the other hand, simulated absorption, even for $L = 5 \mu$m, falls well short of explaining the $\sim10\%$ reflectivity observed for $3 < I_0 < 5\times10^{18}$ W/cm$^2$ from spatially-resolved measurements (Fig.~\ref{figure_PM_reflectplot}, red circles and crosses).   We must therefore attribute the latter result to defocusing of the reflected light out of the $f/12$ microscope collection cone.


\subsection{PM focusing}

\subsubsection{Focusing mechanisms}
A PM irradiated at relativistic intensity focuses the reflected pulse via three mechanisms.  First, the incident pulse induces relativistic transparency by wiggling target electrons at near light speed $c$, causing a decrease $\omega_p^2\to \omega_p^2/\gamma$ in the effective plasma frequency, where $\gamma=\sqrt{1+a_0^2}$. Consequently the plasma's reflecting critical surface dents, independently of the electron density profile itself.  Energy stored temporarily in transverse electron quiver motion can return to the reflected pulse, resulting in focusing without absorption.  Second, ponderomotive pressure of the incident pulse pushes target electrons inward until balanced by outward thermal pressure, denting the electron density profile.   Energy expended in moving these electrons is not returned to the reflected pulse, and is thus absorbed. Third, the pulse can relativistically self-focus in the under-dense pre-plasma without absorption.  Absorption observed for $I_0 \agt 5\times 10^{17}$ W/cm$^2$ ($a_0 \agt 0.5$) indicates that density denting contributes significantly to focusing at these intensities.  However, all three mechanisms likely contribute.  In the Supplementary Material, we present an example of simulated focusing by a PM with $L = 2 \mu$m --- ten times the estimated experimental $L$ at this intensity (see Sec. II) --- irradiated at $I_0 = 10^{17}$ W/cm$^2$, conditions for which absorption and density denting are negligible.  The simulated focal length ($f \approx 100\mu$m) agrees with an analytic estimate of focusing due to relativistic transparency alone, illustrating PM focusing by a single dominant mechanism.  In the following simulations, however, we do not attempt to distinguish the relative contributions of different focusing mechanisms.

\normalsize
\subsubsection{Simulated focusing: $\geq 10^{18}$ W/cm$^2$ PM excitation}
\begin{figure}[!htp]
\centering
\includegraphics[trim={4cm 5cm 0.5cm 5.6cm},width=0.65\textwidth]{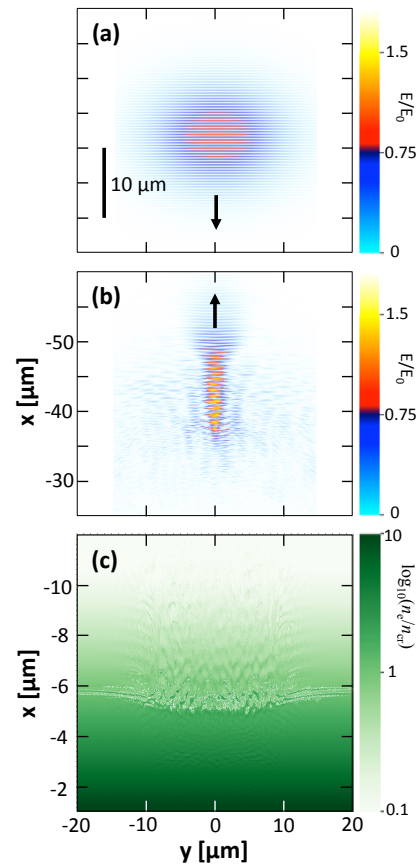}
\caption{2D PIC simulations of (a) incident pulse of peak intensity $I_0 = 3\times10^{18}$, peak field $E_0$, $w_0 = 10 \mu$m, (b) reflected pulse at focus ($x \approx -40 \mu$m) and (c) PM electron density profile at reflection.   Pre-plasma length was $L=2$ $\mu$m.  The normalized peak focused reflected field is $(E/E_0)_{\rm max} = 1.75$. }
\label{focus_1819}
\end{figure}

Figure \ref{focus_1819} shows a 2D PIC simulation of (a) a pulse incident on a PM with $I_0 = 3 \times 10^{18}$ W/cm$^{2}$, $w_0 = 10 \mu$m, $\tau_{\rm FWHM} = 30$ fs, (b) the reflected pulse profile at its focus ($x \approx -40 \mu$m), defined as the point where its peak electric field (intensity) reached a maximum value $E_{max} = 1.76E_0$ ($I_{max} = 3.1 I_0$), and (c) the PM electron density profile at the moment of reflection, which was noticeably dented.  The PM reflected $62\%$ of the incident pulse energy, consistent with the solid blue curve and black square data points in Fig.~\ref{figure_PM_reflectplot}.  The reflected pulse focused to waist $0.08w_0$, duration $1.2\tau_{\rm FWHM}$.


\begin{figure}[t]
\centering
\includegraphics[trim={0 0 0 0cm},width=0.35\textwidth]{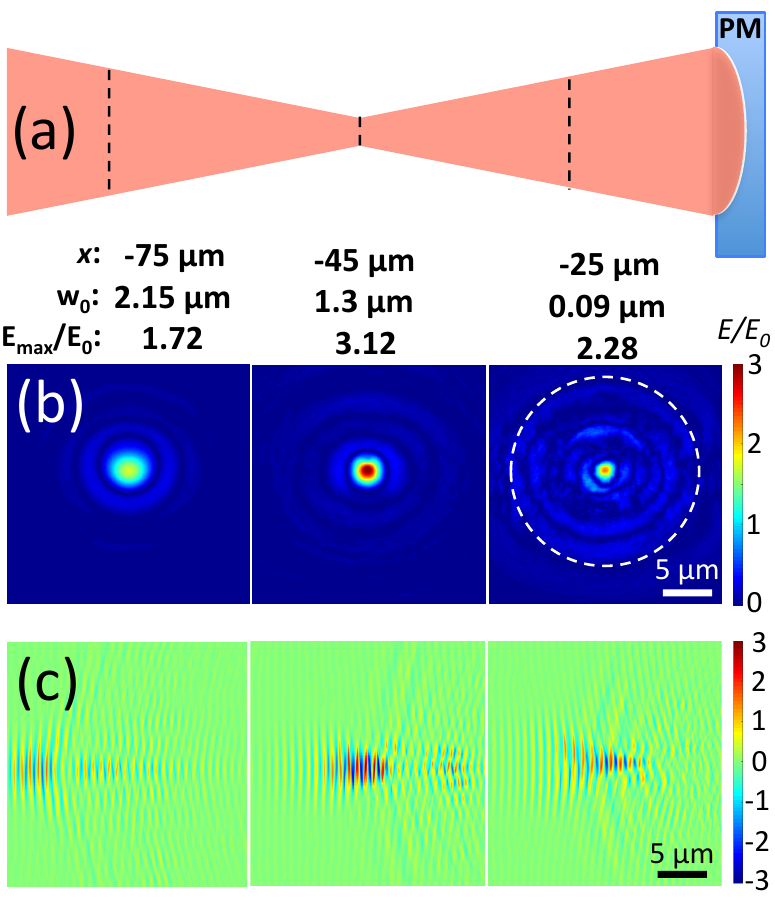}
\caption{3D PIC simulation of normalized electric field amplitude $E(x,y,z)/E_0$ of pulse reflected from PM with pre-plasma length $L = 2 \mu$m. Incident pulse parameters:  $w_0 = 10 \mu$m (dashed white circle in (b)), $I_0 = 3\times10^{18}$ W/cm$^{2}$.  (a) Schematic focused profile of reflected pulse, showing three locations (dashed lines) for which field profiles are displayed;  (b) transverse $E(y,z)/E_0$ and (c) longitudinal $E(y,x)/E_0$ field profiles at each of three distances $x$ from the PM.}     
\label{sim3D}
\end{figure}

This 2D simulation underestimated focused intensity of the reflected pulse because light converged to a line instead of a point.  To correct this, we carried out a 3D simulation (Fig.~9) in which a pulse with transverse profile $a_0 e^{-(y^2+z^2)/w_0^2}$ impinged on the PM with the same incident pulse ($w_0 = 10 \mu$m, $\tau_{\rm FWHM} = 30$ fs, $I_0 = 3\times10^{18}$ W/cm$^2$ ) and pre-plasma ($L = 2 \mu$m) parameters as in the 2D simulation.   The PM reflected $60 \%$ of the pulse energy.   Although the smallest beam waist ($0.09w_0$) was observed at $x \approx -25 \mu$m (Fig.~\ref{sim3D}b, right panel), this spot contained only a small fraction of the reflected pulse energy.  The highest focused field ($E_{max}/E_0 = 3.12$) and highest intensity ($9.7I_0$) were observed at $x \approx -45 \mu$m (Fig.~\ref{sim3D}b, center panel), where the beam waist was $0.13w_0$.  This example shows that 2D and 3D simulations yielded similar $R$ and $f$, but that 2D simulations underestimated focused field strength (intensity) by approximately a factor of 1.8 (3.2).  In addition, the 3D simulation yielded a $50\%$ larger focal spot and a more longitudinally extended focal region, indications that the PM's spherical aberration was more pronounced in the 3D simulation.  


We performed 2D simulations of the reflectivity of a 400 nm \emph{probe} pulse at time delays $-0.25 < \Delta t < 1.25$ ps after 800 nm pump excitation of the PM at $5 \times 10^{18}$ W/cm$^2$, assuming $L = 2 \mu$m.  Filled circles in Fig.~\ref{figure_PM_prbplot}(b) show the portion of the reflected probe pulse energy contained within a $45$ mrad cone angle, mimicking the angle of the probe collection lens.  For comparison, the green triangles show the \emph{total} probe energy reflected at all angles.  Whereas the latter remains $> 0.9$ throughout the simulated interval, the simulated restricted-angle reflectivity qualitatively mimics observed reflectivity dynamics:  an initial drop from 0.97 ($\Delta t = 0$) to 0.75 ($\Delta t = 0.5$ ps), followed by partial recovery to 0.85 ($\Delta t = 0.75$ ps), then a steep drop at later times.  

Fig.~\ref{figure_PM_prbplot}(c) conceptually explains observed and simulated probe reflectivity dynamical trends.  Initially ($\Delta t \alt 0$) overdense plasma forms over the entire probe focal spot, with minimal curvature of the reflecting surface.  Thus the probe reflects nearly totally into an $f/12$ probe collection cone.  As the reflecting surface dents ($0 < \Delta t < 0.5$ ps), although total probe reflectivity remains high, a significant fraction of the probe energy focuses tightly in the near field, and thus diverges  beyond the probe collection lens in the far field.  Because the simulation is 2D, the simulated reflectivity drop is less pronounced than in the 3D experiment.  The partial recovery observed for $0.5 < \Delta t < 1$ ps can be attributed to widening and shallowing of the curved reflecting surface, decreasing far-field probe divergence, while total probe reflectivity remains high.  The final drop in probe reflectivity for $\Delta t > 1$ ps can be attributed in part to increased absorption as the pre-plasma lengthens, and in part to increased scatter as the critical surface ripples.


\begin{figure}[htb]
\centering
\includegraphics[trim={0 0 0 0.2cm},width=0.23\textwidth]{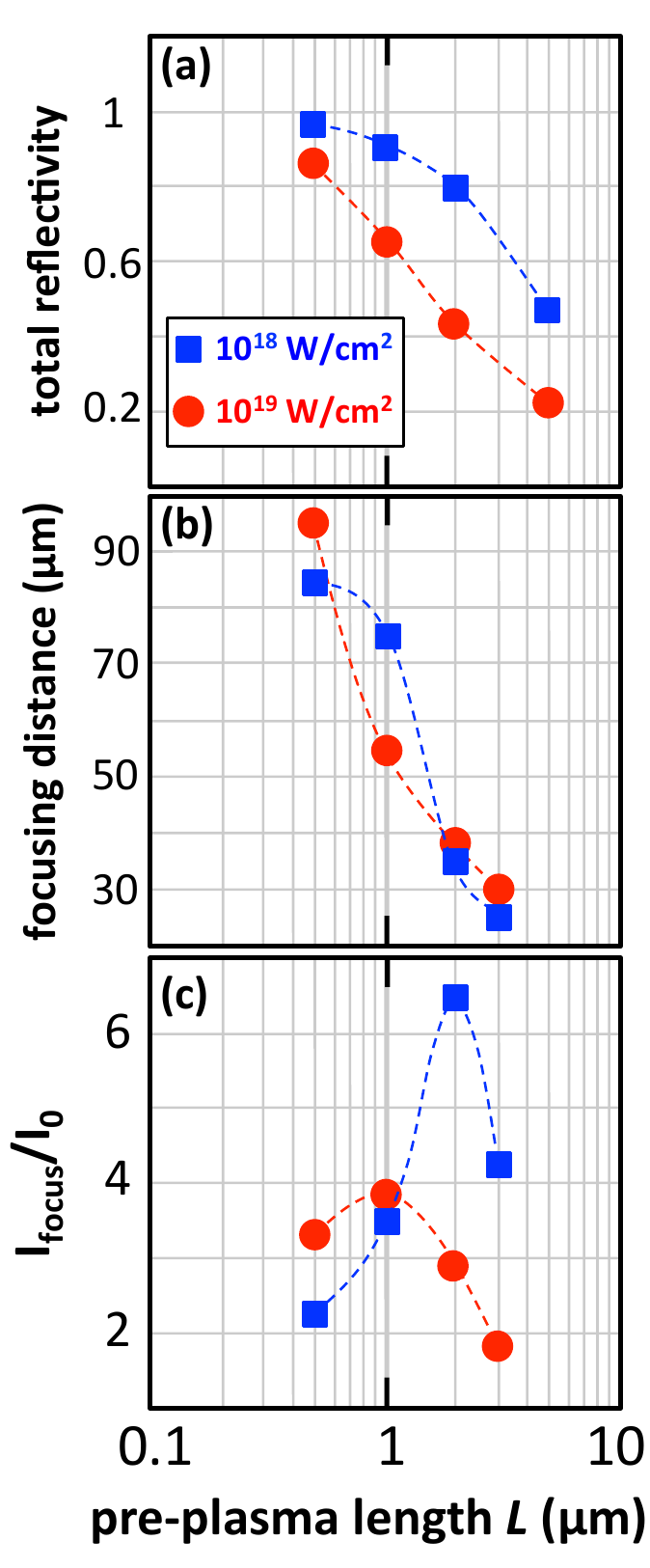}
\caption{Variation with pre-plasma length $L$ of PM optical properties:  (a) reflectivity, (b) focal length $f$ and (c)  intensity enhancement $I_{focus}/I_0$ at focus, for incident intensity $10^{18}$ (blue squares) and $10^{19}$ W/cm$^2$ (red circles), derived from 2D PIC simulations.  Dashed curves are guides to the eye.  $I_{focus}/I_0$ values in (c) underestimate 3D values by $\sim 3\times$. }
\label{figure_PM_RnD}
\end{figure}

\subsubsection{Preplasma scaling of PM optical properties}
So far we have discussed PM optical properties only for pre-plasma length $L = 2 \mu$m.  
Here we vary $L$ in simulations to validate the optimum $L$ found experimentally, and to predict a wide range of relativistic PM properties to be expected in future experiments. 

Fig.~\ref{figure_PM_RnD} plots 2D PIC simulation results for (a) angle-integrated reflectivity, (b) focal length $f$, and (c) intensity increase $I_{focus}/I_0$ at focus vs.~$L$  ($0.5 \le L \le 3 \mu$m) for PMs excited at 
$I_0 = 10^{18}$ (blue squares) and $10^{19}$ W/cm$^2$ (red circles). 
At both $I_0$, total reflectivity drops steeply for $L \agt 0.5 \mu$m  [Fig.~\ref{figure_PM_RnD}(a)] as the incident pulse expends more and more energy denting the surface.  
Simultaneously $f$ decreases monotonically from $f \sim 90 \mu$m at $L = 0.5 \mu$m to $f \sim 25 \mu$m at $L = 3 \mu$m [Fig.~\ref{figure_PM_RnD}(b)].
Accompanying this tighter focus, $I_{focus}/I_0$ grows up to $L = 2 (1) \mu$m at  $10^{18}$ ($10^{19}$) W/cm$^2$ [Fig.~\ref{figure_PM_RnD}(c)].  At larger $L$, however, $I_{focus}/I_0$ drops sharply.  Increasing absorption, which is responsible for the drop in integrated reflectivity for $L>1 \mu$m [see panel (a)], accounts for much of this drop.  Aberration and scatter of the reflected beam, which become pronounced for $L\geq2 \mu$m, accounts for the remainder.  
 The results in Fig.~\ref{figure_PM_RnD} confirm that $1 < L < 2\mu$m is an optimum pre-plasma scale length for relativistic PMs in experiments aimed at maximizing focused intensity close to the PM.  3D $I_{focus}/I_0$ values exceed those plotted in Fig.~\ref{figure_PM_RnD}(c) by a factor of $\sim 3$, as discussed in the previous section.  Plotted $f$ and reflectivity values, however, accurately represent 3D values.  Thus, with an optimized relativistic PM, $I_{focus}$ can exceed $10I_0$ at distances as close as $\sim 25 \mu$m from the PM.  With shorter $L$, Focal lengths as large as $\sim60 \mu$m can be achieved with only two-fold smaller $I_{focus}/I_0$.  Focal lengths shorter than $\sim25 \mu$m, however, appear unachievable without severe losses.  

\subsubsection{Application of focusing PM to Compton backscatter}
In Ref.~11, we speculated that a PM placed at the exit of a LPA could focus a spent LPA drive pulse of $a_0 > 1$ onto trailing electrons, if PM focal length $f \alt R_b$.  Here $R_b$ is the plasma bubble radius, which for stable self-guided propagation is\cite{Lu07} $R_b \approx \lambda_p\sqrt{a_0}/\pi$, where $\lambda_p$ is the plasma wavelength.  Since electrons inject into the back of a bubble ($x = 2R_b$ behind the drive pulse) and accelerate toward its center ($x = R_b$), backscatter occurs in the range $R_b/2 < x < R_b$, depending on how closely accelerating electrons approach the dephasing limit.  PM retro-reflection is self-aligning and efficiently re-cycles the LPA drive pulse,\cite{Phuoc12,Tsai15} avoiding the need for a separate intense backscatter pulse. \emph{Focusing} the retro-reflected drive pulse onto the electrons opens the additional possibility of backscattering at $a_0 >> 1$ without the challenges of aligning and synchronizing a separate tightly focused laser pulse, opening the \emph{nonlinear} Compton regime.\cite{Sarri14}  In this regime, Compton x-rays ($\gamma$-rays) are brighter and higher in photon energy ($E_\gamma \approx 4\gamma_e^2 h\nu_la_0$) than for $a_0 < 1$ ($E_\gamma \approx 4\gamma_e^2 h\nu_l$),\cite{Sarri14}  where $\gamma_e$ is the electron Lorentz factor, $\nu_l$ the laser frequency.  For the terawatt-laser-driven, high-density ($n_e > 10^{19}$ cm$^{-3}$) accelerators studied in Refs.\cite{Tsai15,Phuoc12}, $a_0 \sim 1$ ($I_0 \sim 2\times10^{18}$ W/cm$^2$) at the LPA exit,\cite{Tsai15} yielding $f \sim 40 \mu$m [see Fig.~\ref{figure_PM_RnD}(b)], whereas $R_b \sim 3 \mu$m $<< f$.  Thus the PM focused the laser pulse well beyond trailing electrons, and intensity enhancement at the point of backscatter was inconsequential.  Recently, however, LPAs that accelerate electrons to several GeV by driving tenuous ($n_e < 10^{18}$ cm$^{-3}$) plasma with 0.3 - 0.6 PW laser pulses have been demonstrated.\cite{Wang13,Leemans14}  In this regime, $f \sim R_b \sim 25 \mu$m is achievable, and the present results motivate brief comment on this possibility. 

As a specific example, the experiment of Wang \textit{et al.}\cite{Wang13} used 0.6 PW, 150 fs, $\lambda = 1.056 \mu$m ($h\nu_l = 1.17$ eV) laser pulses focused at $f/50$ into a He-filled cell, creating a $10$ cm plasma column of density $n_e \approx 5\times10^{17}$ cm$^{-3}$.  These pulses blew out plasma bubbles in which $\sim 60$ pC self-injected electron bunches accelerated quasi-monoenergetically to 2 GeV ($\gamma_e \approx 4000$).  The pulses emerged from the cell through a mm exit aperture with $w_0 \approx R_b \approx 25 \mu$m and $a_0 \sim 3$ ($I_0 \sim 10^{19}$ W/cm$^2$).  The 2 GeV electrons exited the cell well before reaching the dephasing limit,\cite{Wang13} and thus propagated near the rear of the bubble at distance $x \approx 2R_b$ behind the laser pulse.  Thus backscatter from a retroreflected drive pulse would occur at $x \approx R_b \approx 25 \mu$m, comparable to the shortest $f$ observed in this work [see Figs.~\ref{figure_PM_2ndfocus}, \ref {focus_1819} and \ref{figure_PM_RnD}(b)].  We thus anticipate backscatter from 2 GeV electrons with $a_0 > 3$, and possibly as large as $\sim 10$, generating bright, directional, fs $\gamma$-ray pulses of photon energy $E_\gamma \approx 75a_0$ MeV for space-time-resolved nuclear science applications using an inexpensive insertion device. 

The studies of Refs.~10,11 angled the PM $\sim 4^\circ$ off normal incidence to avoid retro-reflecting directly into the laser amplifier chain.  To take full advantage of PM focusing, however, the spent pulse must retro-reflect at $\alt \lambda/2f$ ($\alt 1^\circ$) from the PM normal. Several features of GeV-class LPAs  may mitigate the dangers of near-direct retro-reflection compared to MeV LPAs.  First, as the results of Sec. III.C show, most retro-reflected light diverges strongly.  Secondly, the mm-diameter plasma cell entrance aperture is $\sim 10$ cm from the PM, compared to $\sim 3$ mm in the MeV LPAs studied in Refs.~10,11, and thus transmits a thousand-fold smaller fraction of back-reflected light.  Finally, the $\sim 10$ cm long plasma column that the LPA drive pulse created defocuses back-reflected light, further protecting the laser system.  If the hazards of small-angle retro-reflection can be managed, focusing relativistic PMs are promising insertion devices for converting GeV LPAs into Compton $\gamma$-ray sources.

\section{Conclusion}

We have presented an experimental-computational study of the optical properties of plasma mirrors (PMs) irradiated directly at near-normal incidence by high contrast 30 fs, $800$nm laser pulses at intensities (fields strengths) in the range $10^{18} \le I_0 \le 10^{19}$ W/cm$^2$ ($0.68 \le a_0 \le 2.15$).   We find that, as long as the intensity of pre-pulses does not exceed $\sim 10^{14}$ W/cm$^2$ sooner than $\sim 20$ ps prior to the peak of the main pulse --- thereby limiting pre-plasma length $L$ to $\sim 2\mu$m --- the relativistic PM reflects $60-80\%$ of incident light, while focusing a significant fraction of reflected light to intensity as high as $\sim 10I_0$ with focal length $f$ as short as $\sim 25 \mu$m.  Inward denting of the target's electron density profile combines with relativistic transparency and self-focusing in the pre-plasma to create a concave focusing mirror.  Time-resolved reflectivity measurements show the relativistic PM's focusing properties evolve on a ps time scale.  EPOCH simulations show that the optical properties of PMs irradiated in the range $10^{18} \le I_0 \le 10^{19}$ W/cm$^2$ depend strongly on $L$, which therefore provides a key parameter for controlling them.   Applications of focusing relativistic PMs include generation of bright directional Compton $\gamma$-rays from GeV laser-plasma accelerators (LPAs) and coupling of LPA stages.

\section*{Supplementary Material}
See Supplementary Material for transverse interferometry measurements of the relativistic PM and simulated PM focusing at excitation intensity $10^{17}$ W/cm$^2$. 

\begin{acknowledgments}

DOE grants DE-SC0012444, DE-SC0011617 and AFOSR grant FA9550-14-1-0045 supported experimental work. 
DOE contracts DE-SC0007889, DE-SC0010622 supported 3D PIC modeling.  AFOSR Contract FA9550-14-1-0045, NNSA Contract DE-FC52-08NA28512, and DOE Contract DE-FG02-04ER54742 supported simulations by A. V. A., which used the EPOCH code (developed under UK EPSRC grants EP/G054940/1, EP/G055165/1 and EP/G056803/1) and resources from the Texas Advanced Computing Center.  A DOE Computational Sciences Graduate Fellowship administered under
Contract DE-AC05-06OR23100 supported D.J.S.

\end{acknowledgments}

\end{document}